\documentclass[notitlepage,superscriptaddress,showpacs,showkeys,nobalancelastpage,twocolumn,aps,longbibliography,prb]{revtex4-2}
\RequirePackage[T1]{fontenc}
\RequirePackage{times} 

\usepackage{siunitx} 
\usepackage[italicdiff]{physics}
\usepackage{amsfonts}
\usepackage{lipsum}
\usepackage{multirow}
\usepackage{subfigure}
\usepackage{amsmath}
\usepackage{amssymb}
\usepackage{graphicx, epstopdf}
\usepackage{wasysym}
\usepackage{xspace}
\usepackage{array}
\usepackage[hidelinks]{hyperref}
\usepackage[dvipsnames]{xcolor}
\usepackage{my_symbols}
\usepackage{CJK}
\usepackage{bm}
\usepackage{verbatim}
\usepackage{tikz}
\usepackage{comment}
\usepackage{marginnote}
\usepackage[textwidth=1.5cm]{todonotes}
\usepackage[normalem]{ulem}
\usepackage{soul}
\usepackage{booktabs}
\usepackage[commandnameprefix=ifneeded]{changes}

\sethlcolor{green}

{\par}

\newcounter{mycomment}

\excludecomment{comment}  

\begin{document}

\begin{CJK*}{UTF8}{gbsn} 
\title{Micromagnetic Study of the Dipolar-Exchange Spin Waves in Antiferromagnetic Thin Films}
\author{Jiongjie Wang (王炯杰)}
\affiliation{Department of Physics and State Key Laboratory of Surface Physics, Fudan University, Shanghai 200433, China}
\author{Jiang Xiao (肖江)}
\email{xiaojiang@fudan.edu.cn}
\affiliation{Department of Physics and State Key Laboratory of Surface Physics, Fudan University, Shanghai 200433, China}
\affiliation{Institute for Nanoelectronics Devices and Quantum Computing, Fudan University, Shanghai 200433, China}
\affiliation{Shanghai Research Center for Quantum Sciences, Shanghai 201315, China}
\affiliation{Hefei National Laboratory, Hefei 230088, China}

\keywords{spin waves, spintronics, antiferromagnetics, magnetic properties of thin films}

\pacs{75.30.Ds, 85.75.-d, 75.50.Ee, 75.70.-i}

\begin{abstract} 
  In antiferromagnets, dipolar coupling is often disregarded due to the cancellation of magnetic moments between the two sublattices, leaving spin-wave dispersion predominantly determined by exchange interactions. However, antiferromagnetic spin waves typically involve a slight misalignment of the magnetic moments on the sublattices, giving rise to a small net magnetization that enables long-range dipolar coupling. In this paper, we investigate the role of this dipolar coupling in spin-wave excitations and its influence on the spin-wave dispersion. Our findings show that: (i) when the N\'{e}el vector is perpendicular to the film plane or lies within the film plane and parallel to the wave vector, the dispersion branches can be divided into two groups—those unaffected by the dipolar field and those influenced by it. In these cases, the total magnetic moment remains linearly polarized, but the polarization directions differ between the two types of branches; (ii) when the N\'{e}el vector lies in the film plane and is perpendicular to the wave vector, the dipolar interactions affect both types of dispersion branches, leading to their hybridization. This hybridization alters the polarization of the magnetic moment, resulting in elliptical polarization.

\end{abstract}

\maketitle
\end{CJK*}

\section{Introduction}

Spin waves 
\cite{gurevichMagnetizationOscillationsWaves1996,stancilSpinWaves2009}, the collective excitations of localized magnetic moments in ordered spin systems, have garnered significant attention for their fundamental importance and potential applications. In both ferromagnetic and antiferromagnetic materials, these magnonic excitations propagate through the medium without involving net charge transport, offering unique opportunities for low-power, highly efficient information processing in spintronic and magnonic devices \cite{chumakMagnonSpintronics2015,yu_magnetic_2021}. Spin waves in ferromagnets and antiferromagnets exhibit distinctive dispersion relations and dynamics due to differences in their magnetic exchange interactions and symmetry properties, enabling a rich spectrum of physical phenomena that arise from their interaction with external fields, coupling to other quasiparticles, or structural inhomogeneities. 

Spin waves in ferromagnetic materials can generally be classified into two types: exchange spin waves and dipolar spin waves \cite{gurevichMagnetizationOscillationsWaves1996,stancilSpinWaves2009}. Exchange spin waves dominate at short wavelengths and are governed by the strong quantum mechanical exchange interaction between neighboring spins. Dipolar spin waves, on the other hand, are prominent at longer wavelengths and are primarily influenced by the long-range dipolar interactions between magnetic moments. 
The comprehensive theory of dipolar-exchange spin waves in ferromagnetic thin films has been well-established by De Wames \etal \cite{dewames_surface_1969} and many others 
\cite{goedsche_spin_1970,kalinikosSpectrumLinearExcitation1981a,mika_dipolar_1985,kalinikosTheoryDipoleexchangeSpin1986a,hillebrands_spinwave_1990,harmsTheoryDipoleexchangeSpin2022}.

Spin waves in antiferromagnetic materials are predominantly governed by exchange interactions due to the alternating alignment of magnetic moments within the lattice \cite{keffer_theory_1952,marshallSpinWaveTheoryAntiferromagnetism1955,oguchi_theory_1960,descloizeauxSpinWaveSpectrumAntiferromagnetic1962,loudon_effect_1963,wolfram_surface_1969,camley_longwavelength_1980,luthi_dipolar_1983,stamps_dipoleexchange_1987,pereira_theory_1999,wieserQuantizedSpinWaves2009,shen_magnon_2020,liuDipolarSpinWaves2020,sun_dipolar_2024}, which cancels the net magnetization at the macroscopic scale. Spin waves in antiferromagnets exhibit a richer spectrum compared to ferromagnetic systems due to the presence of two spin sublattices, often leading to both acoustic and optical modes in their dispersion relations, as well as the additional polarization degree of freedom \cite{chengAntiferromagneticSpinWave2016,lan_antiferromagnetic_2017,proskurinSpinWaveChiralityIts2017,yu_polarization-selective_2018} that ferromagnetic spin wave doesn't have. Dipolar spin waves are relatively less studied in antiferromagnetic materials because of the inherent cancellation of net magnetization, which minimizes the influence of long-range dipolar interactions. However, such cancellation does not hold when there are antiferromagnetic excitations, for which a small net magnetization arises, which then gives rise to dipolar coupling. 


Micromagnetic simulations \cite{abertMicromagneticsSpintronicsModels2019} have become a cornerstone in the study of spin waves within magnetic materials, offering deep insights into their dynamics and interactions. While simulations for ferromagnetic spin waves are well-established due to the simpler single-sublattice dynamics and widespread applicability in magnonic devices, the micromagnetic simulation for antiferromagnetic spin waves remains comparatively underdeveloped. This is largely attributed to the intrinsic challenges posed by their dual-sublattice structure and ultrafast precession frequencies, which demand advanced computational techniques. Nevertheless, recent innovations in numerical methods and the adaptation of simulation frameworks, such as COMSOL Multiphysics \cite{_comsol_}, have made significant strides in overcoming these limitations. Specifically, the development of \textit{ms-comsol} module \cite{yu_comsol_nodate}
in COMSOL Multiphysics facilitates the exploration of dipolar-exchange spin waves in antiferromagnetic systems. 

This paper presents a new approach to simulating dipolar-exchange spin waves in antiferromagnetic thin films. We demonstrate the efficacy of combining the \textit{ms-comsol} module, used for micromagnetic simulation, with the \textit{mfnc} module, responsible for accurate calculation of dipolar fields. This combined methodology allows for a comprehensive simulation capturing the intricate interplay of dipolar and exchange interactions in these systems. We also analyze the dipolar-exchange spin wave in antiferromagnetic thin films, and found that the dipolar field affect the spin wave dispersions depending on their polarizations in antiferromagnet thin film. Our results provide valuable insights into the dynamic behavior, especially the polarization, of spin waves in antiferromagnetic materials, offering a foundation for future research in magnonics and spintronics.

\section{Micromagnetic Simulation of Antiferromagnetic Spin Wave}

The antiferromagnetic spin waves can be described by two coupled Landau-Lifshitz-Gilbert equations:
\cite{keffer_theory_1952}
\begin{subequations}
  \label{eqn:LLGmm}
  \begin{align}
    \dot{\mb_1}(\br,t) &= -\gamma \mb_1\times(\bH_1 + \bh) + \alpha \mb_1\times \dot{\mb_1} \\
    \dot{\mb_2}(\br,t) &= -\gamma \mb_2\times(\bH_2 + \bh) + \alpha \mb_2\times \dot{\mb_2},
  \end{align}
\end{subequations}
where $\bH_i = -\delta F(\mb_1, \mb_2)/\delta \mb_i$ is the effective field acting on sublattice $i$ with free energy functional \cite{yu_magnetic_2021}: 
\begin{equation}
  F(\mb_1, \mb_2) 
  = \sum_i\int d^3\br(f_i^Z + f_i^A + f_i^{E}) + \int d^3\br f^{E}, 
\end{equation}
where $f_i^{Z, A, E}$ are the Zeeman, anisotropy, and intra-sublattice exchange energies density for sublattice $i$, and $f^{E}$ is the inter-sublattice exchange energy density between the two sublattices: 
\begin{align}
  f_i^Z &= -\mu_0M_s\mb_i\cdot \qty(\bH_\text{ext} + \bh), \nn
  f_i^A &= - \frac{K}{2} (\mb_i\cdot\hat{\bz})^2, \nn
  f_i^{E} &= \half A'(\nabla \mb_i)^2, \nn
  f^{E} &= J\mb_1\cdot \mb_2 - A(\nabla \mb_1)\cdot(\nabla \mb_2). \nonumber
\end{align}
Here the easy uniaxial anisotropy is along $\hbz$ with strength $K$, thus the equilibrium magnetic moments and the N\'{e}eel vector, are assumed to point in $\hbz$ direction. $A = Ja^2/2$ and $A'=J'a^2/2$ represent the inter- and intra-sublattice exchange coupling (Appendix \ref{app:exchange}).
And $\bh = \bh(\mb)$ is the dipolar field due to the net magnetic moment distribution $\mb = \mb_1+\mb_2$. There are two assumptions for the dipolar fields: first the dipolar fields are generated by the net moment $\mb$ and we do not consider difference in dipolar field from $\mb_1$ and $\mb_2$ due to the spatial separation on the atomic scale; the dipolar field on the two sublattice are the same, also because of the negligible atomic scale of their separations. 

By recombining the two sublattices into a net magnetic moment $\mb = \mb_1 + \mb_2$ and a N\'eel vector $\bn = \mb_1 - \mb_2$, we can rewrite \Eq{eqn:LLGmm} as the equivalent equations of motion for $\mb$ and $\bn$ (Eq.(B1) in Appendix \ref{app:semi-analytical}). When the $J\mb_1\cdot\mb_2$ is the dominating exchange coupling, which is true for most natural antiferromagnet, the roles of intralayer exchange $A'$ and interlayer exchange $A$ are interchangeable. To simplify the micromagnetic simulation, we use $A'$ to account for the roles of both $A$ and $A'$ because in this way we can couple two independent LLG simulations for \Eq{eqn:LLGmm} and only communicate $\mb_1$ and $\mb_2$ between the two simulations, and no need to communicate their gradients $\nabla\mb_1$ and $\nabla\mb_2$.

\begin{figure}[t]
  \centering
  \includegraphics[width=\columnwidth]{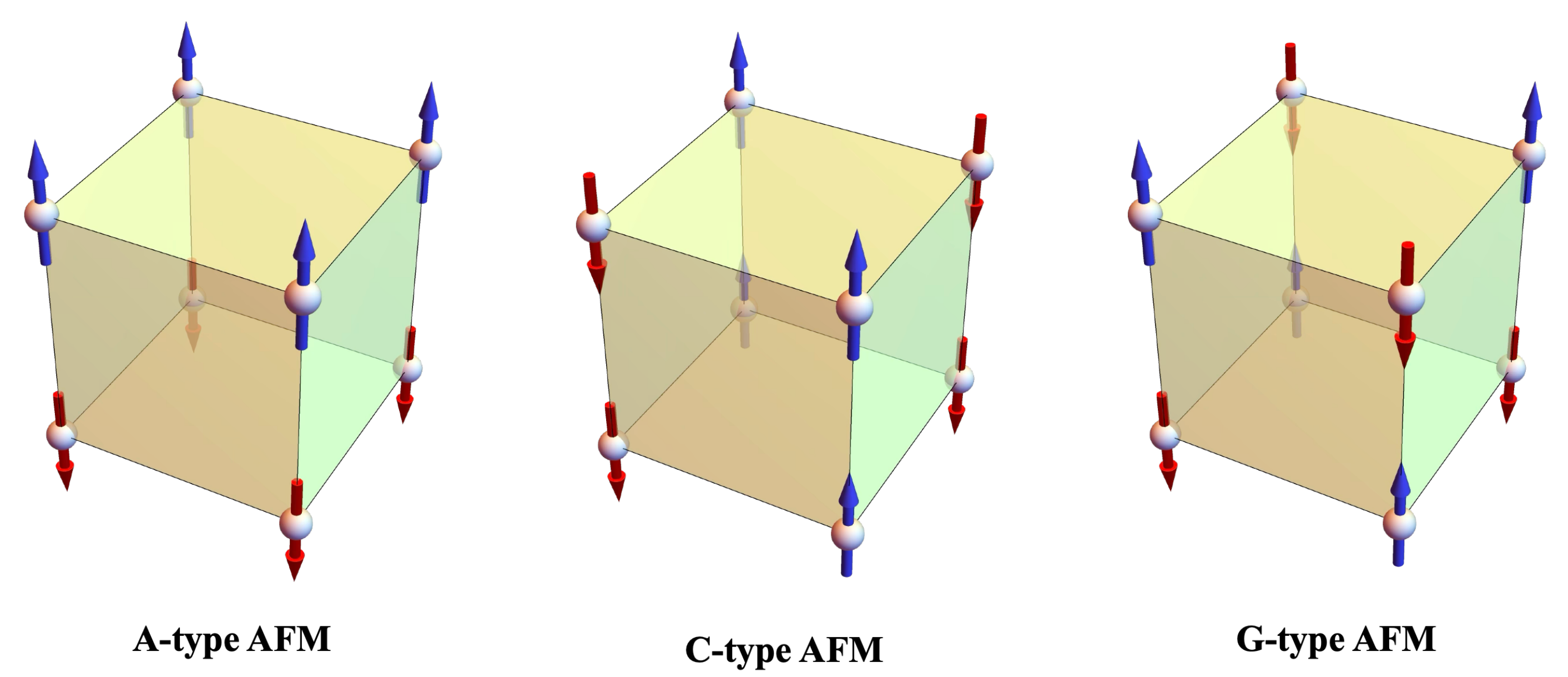}
  \caption{Three types of antiferromagnet with cubic lattice: A-type, C-type, and G-type.}
  \label{fig:AF_ACG}
\end{figure}

There are mainly three types of antiferromagnet, the A-type, C-type, and G-type \cite{solovyev_magnetooptical_1997,treves_magnetic_1962},
 as depicted in \Figure{fig:AF_ACG}. For simplicity, we assume cubic lattice with equal lattice constants in three directions $a = b = c$. 
If we only consider nearest-neighbor coupling and ignore the next-nearest-neighbor coupling, then the parametrization for these three types of AF correspond to: 
i) A-type, $\bA' = A'(1, 1, 0), \bA = A(0, 0, 1)$; 
ii) C-type, $\bA' = A'(0, 0, 1), \bA = A(1, 1, 0)$; 
iii) G-type, $\bA' = 0, \bA = A(1, 1, 1)$. 
For simplicity in our discussion, we will focus subsequently on the G-type antiferromagnets, where the exchange interaction remains isotropic.

In antiferromagnets, spin waves are primarily governed by exchange interactions, often leading to the neglect of dipolar fields due to the cancellation of net magnetization from opposing sublattices. However, this cancellation is disrupted in the presence of antiferromagnetic excitations, which generate a net magnetic moment as demonstrated by Keffer and Kittel \cite{keffer_theory_1952}. Consequently, these excitations introduce non-vanishing dipolar fields that may influence the spin wave dispersion. 


We carry out the full micromagnetic simulations using the {\it ms-comsol} module, a micromagnetic simulation package based that we developed on COMSOL Multiphysics \cite{_comsol_} in the past few years \cite{lan_spin-wave_2015,yu_magnetic_2016,yu_polarization-selective_2018,yu_magnetic_2020, yu_hopfield_2021,zhang_frequency-domain_2023}. The module was originally developed for simulating ferromagnetic spin waves, and later extended into simulating spin waves in synthetic antiferromagnets \cite{lan_antiferromagnetic_2017}. In this work, we further extend the capability of this module to include the influence of dipolar fields in antiferromagnetic spin waves. The simulations for antiferromagnet are carried out by treating the antiferromagnetic dynamics as two coupled ferromagnetic simulations for two sublattices based on \Eq{eqn:LLGmm} using the {\it ms-comsol} module. The dipolar field is included using the {\it mfnc} module (magnetic fields, no current) built-in in COMSOL, which is solved via the magnetic scalar potential. The two modules are coupled in COMSOL by exchanging their outputs: the net magnetic moment $\mb(\br,t) = \mb_1+\mb_2$ calculated from {\it ms-comsol} is fed into the {\it mfnc} module to compute the magnetic scalar potential in quasi-static limit, thereby the dipolar field $\bh(\br,t)$; this dipolar field from the {\it mfnc} module is then fed back into the {\it ms-comsol} module to simulation the magnetization dynamics. Through the self-consistent iterations between the {\it ms-comsol} module and the {\it nfnc} module, we may simulate the antiferromagnetic resonance with dipolar fields taken into account.

\begin{figure}[t]
  \centering
  \includegraphics[width=0.8\columnwidth]{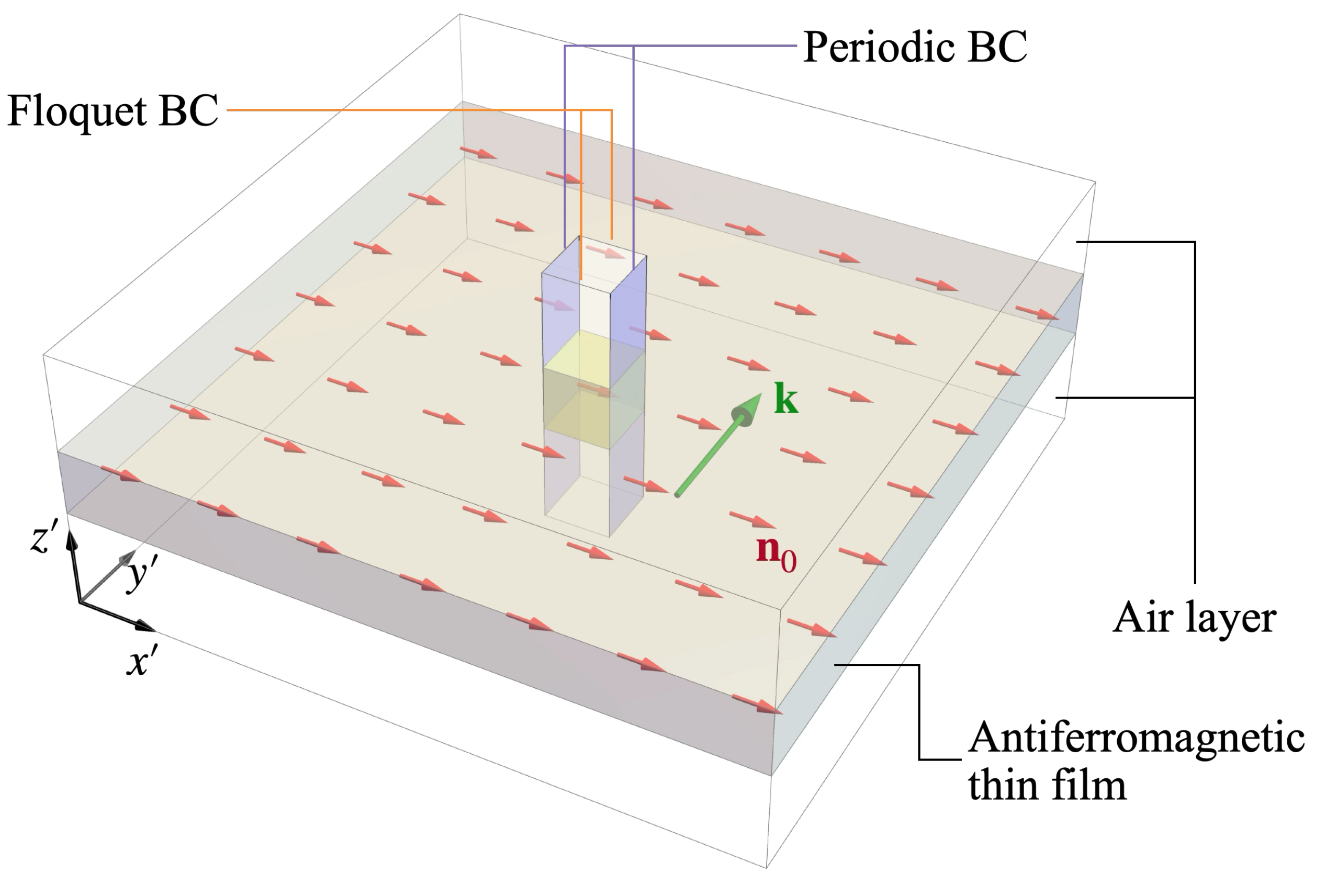}
  \caption{Schematic diagram of the geometry and boundary conditions used in the micromagnetic simulation. The computational space includes the magnetic region for solving the LLG equation and the entire space for solving the magnetic Gauss's law. Due to the translational invariance of the infinitely large thin film, we only need to simulate a small pillar region, where we use periodic boundary condition in the direction perpendicular to the wave vector and  Floquet boundary condition in the direction parallel to the wave vector. Here $x', y', z'$ are the coordinate system of the thin film with $\hat{z}'$ pointing in the film normal.}
  \label{fig:COMSOL_model}
\end{figure}

The system we focus in this paper is an infinite antiferromagnetic thin film. To incorporate the dipolar fields in {\it mfnc} module, we need to buffer the thin film with air layer above and beneath the thin film (see \Figure{fig:COMSOL_model}). Period boundary condition is used in direction perpendicular to the in-plane wave vector, and Bloch-Floquet periodic boundary condition is used in the direction of the wavevector \cite{vandenboom_fully_2021}.
Consequently, only a one-dimensional strip along the thickness direction is required for simulation (the pillar region in \Figure{fig:COMSOL_model}), greatly reducing the simulation complexity and time. 
To reduce the computation even further, we adopt the frequency-domain simulation for \emph{ms-comsol} + \emph{mfnc} in COMSOL , which eliminates the temporal simulation completely. The details about the frequency-domain micromagnetic simulation using \emph{ms-comsol} can be found in Ref. \cite{zhang_frequency-domain_2023}.

\section{Semi-analytical Approach}

For the special case of infinite (anti-)ferromagnetic thin films focused in this paper, we can also employ a semi-analytical approach to calculate the spin wave dispersion. This method was developed by De Wames and Wolfram in the 1960s \cite{dewames_surface_1969} in ferromagnets and was applied to antiferromagnetic systems in the 1980s \cite{stamps_dipoleexchange_1987}.

The basic idea is to convert the LLG equation into a susceptibility $\mb$ and $\bn$ in response to the dipolar fields $\bh$ in frequency and reciprocal space domain. On the other hand, the dipolar fields $\bh$ and $\mb$ satisfy the Maxwell equations in quasi-static approximation. By expressing the dipolar fields in terms of an unknown magnetic scalar potential $\bh = -\nabla \phi$, the combined formulation of coupled LLG equations and Maxwell equations are converted into a 6-th order ordinary differential equation for the scalar potential $\phi$ based on the magnetic boundary conditions are the interfaces of the antiferromagnetic thin film and air. Eventually, the dispersion relation can be obtained by solving the eigenvalue problem of the 6-th order ordinary differential equation (see Appendix \ref{app:semi-analytical} for the detailed formulation).

The semi-analytical approach proves highly effective for homogeneous systems, such as infinite thin films, making it a valuable tool for benchmarking micromagnetic simulations and examining dipolar-exchange spin wave physics. However, its applicability diminishes when addressing non-trivial geometries or systems with inhomogeneous structures. In contrast, micromagnetic simulations demonstrated here offer greater flexibility and are well-suited for analyzing complex systems, including finite sizes, intricate geometries, and diverse magnetic textures.

\begin{center}
  \begin{table}[b]
    \centering
    \begin{tabular}{cccl}
      \toprule
      Parameter & Value & Unit & Description \\
      \midrule
      $\omega_\ssf{M}$ & 34.5 & \si{GHz} & Saturation magnetization \\
      $K$ & $0.25 \omega_\ssf{M}$ & \si{GHz} & anisotropy \\
      $J$ & $80 \omega_\ssf{M}$ & \si{GHz} & inter-sublattice exchange \\
      $A$ & $4.74\times 10^{-3}$ & \si{GHz\cdot\mu m^2} & inter-sublattice exchange \\
      $A'$ & 0 & \si{GHz\cdot\mu m^2} & intra-sublattice exchange \\
      $d$ & 3.05 & \si{\mu m} & film thickness \\
      $d_{air}$ & 35.075 & \si{\mu m} & upper/lower air thickness \\
      \bottomrule
    \end{tabular}
    \caption{Parameters used in the calculation. \cite{yan_allmagnonic_2011,lan_antiferromagnetic_2017}}
    \label{tab:para}
  \end{table}
\end{center}

\section{Results}

\begin{figure*}[ht]
  \centering
  \includegraphics[width=\textwidth]{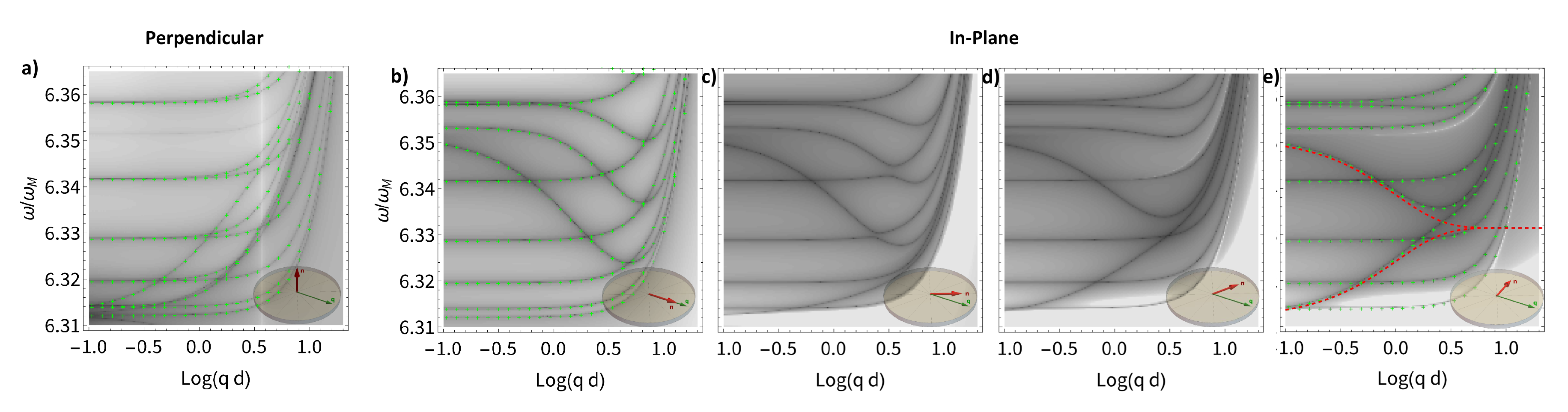}
  \caption{The results from the semi-analytical calculation (gray-scale) and {\it ms-comsol} simulations (green crosses) for the perpendicular (a) and in-plane (b-e) magnetized thin film. The red and green arrow denote the direction of the N\'{e}el vector and the wave vector, respectively. For the in-plane case, the angle between the two vectors are $0, \pi/6, \pi/3, \pi/2$, respectively. The red curves in the rightmost panel correspond to dispersions for the dipolar spin surface wave modes in the long-wavelength limit (given by Eq.(C4) in Appendix \ref{app:surface_wave}).}
  \label{fig:matlab}
\end{figure*}

We are now ready to  present the results for the dipolar-exchange antiferromagnetic spin waves. We first validate the {\it ms-comsol} simulation by comparing the results with the semi-analytical approach. Then we present the dispersions and the spin wave profiles for the perpendicularly magnetized and in-plane magnetized antiferromagnetic thin films.
The parameters used in the COMSOL simulation and semi-analytical calculation are summarized in \Table{tab:para}. 

We examine two distinct configurations concerning the orientation of the Ne\'el vector relative to the film: i) the perpendicular case, where the Ne\'el vector is perpendicular to the film plane (i.e., \(\bn_0 \parallel \hbz = \hbz'\)), and ii) the in-plane case, where the vector lies within the plane of the film. In the first scenario, the structural configuration possesses rotational symmetry about the axis defined by \(\hbz\), resulting in a dispersion relation that is solely a function of the magnitude of the in-plane wave vector, independent of its directional orientation. Conversely, in case ii), the absence of rotational symmetry about the film's normal leads to a dispersion relation that is dependent on both the magnitude and the direction of the in-plane wave vector, articulated through the angle \(\theta\) between \(\bn_0\) and \(\bq\). 

\subsection{Validation of \emph{ms-comsol} Simulation with Dipolar fields}

\Figure{fig:matlab} presents the dispersion relations obtained by micromagnetic simulation using {\it ms-comsol} (the green crosses) and the semi-analytical approach (the gray-scale color map) for the perpendicular and in-plane scenarios. 
The gray-scale color map further illustrates the determinant of the \(M\) matrix in Appendix \ref{app:semi-analytical} ($-\sqrt{-\log\abs{M(\omega,\bq)}}$), where local minima signify conditions under which the determinant vanishes, thereby indicating eigenfrequencies for the spin wave excitations.
The green crosses are the results from the {\it ms-comsol} using the eigenvalue solver in COMSOL Multiphysics. 
The results are in excellent agreement, confirming the reliability of the {\it ms-comsol} simulation.

\subsection{Dispersions and Profiles}

\begin{figure*}[t]
  \centering
  \includegraphics[width=\textwidth]{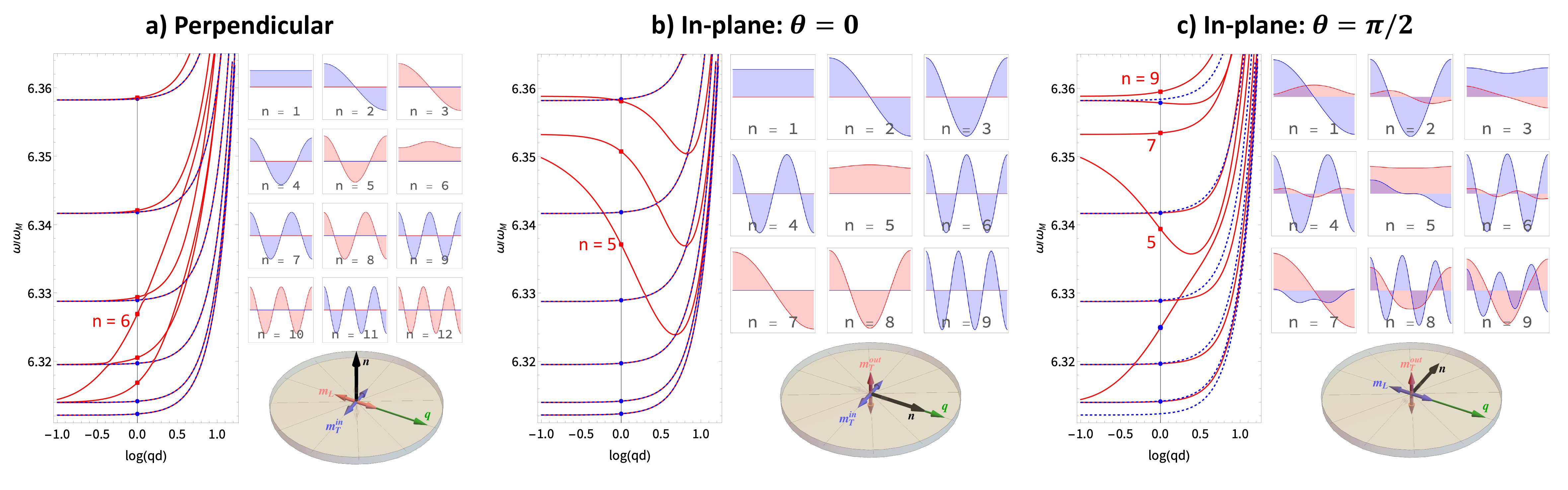}
  \caption{The spin wave dispersion and the spin wave profiles for the net magnetization for the perpendicularly magnetized (a) and in-plane (b, c) magnetized antiferromagnetic thin films. 
  The blue and red components in the profiles correspond to the net magnetization $\mb$ projected onto the two transverse directions indicated in the lower vector diagrams. Here $n$ is the index for the branch from low frequency to high frequency.
  }
  \label{fig:disp}
\end{figure*}


We now analyze the antiferromagnetic spin wave in more detail by examining the behaviors of the dispersion relations and the corresponding spin wave profiles. Since the results calculated from the semi-analytical approach and simulated from COMSOL are identical as demonstrated in \Figure{fig:matlab}, in the following we only show the results calculated from COMSOL simulation ({\it ms-comsol} + {\it mfnc}). \Figure{fig:disp} shows the more detailed dispersions and corresponding spin wave profiles for the perpendicular case and the in-plane case (for $\theta = 0, \pi/2$). 

To accurately characterize the polarization of spin wave excitations, we decompose the net magnetization (\(\mb = \mb_1 + \mb_2\)) into three distinct components based on their oscillation directions relative to the wave vector \(\bq\). This results in two types of polarization akin to phonon polarizations: one longitudinal polarization, where the magnetization oscillates along \(\bq\), and two transverse polarizations, where the magnetization oscillates perpendicularly to \(\bq\). Given that \(\bq\) lies within the film plane, longitudinal polarization is confined to in-plane oscillations, while two perpendicular directions yield in-plane and out-of-plane transverse polarizations. In antiferromagnetic systems, only two polarization states can coexist due to the constraint of magnetization orientation being perpendicular to the N\'{e}el vector \(\bn_0\), effectively eliminating one possible direction and leaving a pair of polarizations that depend on the orientation of \(\bn_0\).

\Figure{fig:disp}(a) shows the dispersion and the typical spin wave profiles for the case where the N\'{e}el vector points perpendicularly to the thin film ($\bn_0 \parallel \hbz'$). 
The blue and red components in the profiles in \Figure{fig:disp}(a) correspond to the net magnetization $\mb$ projected onto the in-plane transverse direction ($\mb\cdot \be_T^{\rm in}$, see lower right panel in \Figure{fig:disp}(a)) and the longitudinal direction ($\mb\cdot\be_L$), respectively.
The profiles in \Figure{fig:disp}(a) show only one of the components is non-zero, meaning that all eigen modes are linearly polarized, either in the in-plane transverse polarization ($n = 1, 2, 4, 7, 9, 11$) or the (in-plane) longitudinal polarization ($n = 3, 5, 6, 8, 10, 12$).
In order to see the effect of dipolar fields, we also calculate the pure exchange spin wave dispersion (dashed blue curves) without the dipolar fields (by turning off the {\it mfnc} module). The spin wave modes with in-plane transverse polarization matches with the exchange-only dispersions, which means that the dipolar interaction does not affect these modes. On the other hand, the modes with in-plane longitudinal polarization deviate from the exchange-only dispersions, especially at large $q$ with $qd\gtrsim 1$, which means that the longitudinal polarization is influenced by the dipolar fields.
This distinction between the two in-plane polarizations (transverse and longitudinal) is of no coincidence. For an infinite thin film, if the magnetization is lying within the plane and uniform, there should be no dipolar fields. This is the case for the in-plane transverse polarization case, where the magnetization points in the transverse direction ($\be_T^{\rm in}$) and uniform in that direction. On the other hand, for the (in-plane) longitudinal polarization, even though the magnetization also lies in-plane (pointing in the $\bq$ direction), the magnetization is not uniform but alternating in sign along $\bq$ direction. This non-uniform magnetization gives rise to the non-vanishing dipole fields for the in-plane longitudinal modes, thus influencing their dispersions.


\Figure{fig:disp}(b,c) shows the results for the two cases with N\'{e}el vector lying in the film plane ($\bn_0 \perp \hbz'$) at $\theta = 0$ (\Figure{fig:disp}(b)) and $\theta = \pi/2$ (\Figure{fig:disp}(c)), \ie the in-plane N\'{e}el vector is parallel ($\bn_0 \parallel \bq$) or perpendicular ($\bn_0 \perp \bq$) to the wave vector.

For the parallel case ($\bn_0 \parallel \bq$), similar to the case with out-of-plane N\'{e}el vector in \Figure{fig:disp}(a), all dispersion branches can also be categorized into two types of modes, being affected or unaffected by the dipolar interaction. Such distinction also arises from the polarization characters of the corresponding modes. The modes with in-plane transverse polarization ($n = 1, 2, 3, 4, 6, 9$) does not produce dipolar fields for the same reason as the above. While, the modes which has out-of-plane transverse polarization ($n = 5, 7, 8$) produce dipolar fields or demagnetization fields, so their dispersions are lifted.   
For finite $q$, the magnetization direction alternates (in $\hbz$ and $-\hbz$) along the direction of $\bq$, and the resulting dipolar fields are weakened, leading to the downward bending in dispersions. This behavior is similar to the backward volume modes in the ferromagnetic thin films \cite{prabhakar_spin_2009}. 

When the wave vector is oriented perpendicular to the in-plane N\'{e}el vector ($\bq \perp \bn_0$), the results are slightly different from the two scenarios discussed above as shown in \Figure{fig:disp}(c). 
Here, all dispersions are influenced by dipolar effects. The orientation of $\bn_0$ indicates that the polarization degree of freedom is now in the longtudinal direction ($\be_L \parallel \bq$) and out-of-plane transverse direction ($\be_T^{\rm out} \parallel \hbz'$). We have seen from above that both the longitudinal and the out-of-plane transverse polarization produce dipolar fields, therefore all dispersions shall observe some dipolar corrections. 
Furthermore, different from the previous two scenarios where all modes are linearly polarized in one of the three possible polarization directions, spin wave profiles in \Figure{fig:disp}(c) show that the eigenmodes in this scenario are not linearly polarized, but have spatially varying elliptical polarization with magnetization components in both the longitudinal ($\be_L$) and (out-of-plane) transverse ($\be_T^{\rm out}$) directions. This indicates that the longitudinal mode and the (out-of-plane) transverse mode are no longer independent, and they couple to one another via their dipolar fields, resulting new hybridized normal modes with more complicated spatially varying elliptical polarizations. 
The modes with significant out-of-plane transverse magnetic components ($n = 5, 7, 9$), manifest more pronounced dipolar effects (starting from $q \to 0$), in line with the strong demagnetization field due to out-of-plane magnetization.
In contrast, the modes with primarily longitudinal features ($n = 1, 2, 4, 6, 8$) exhibit dipolar effects only at finite wave vectors ($qd \gtrsim 1$) for a similar reason as the case in \Figure{fig:disp}(a).  
Notably, the longitudinal mode for $n = 3$ stands out with significantly enhanced dipolar effects, which can be attributed to its nearly uniform magnetization through the thickness, reducing field cancellation from different layers.

\section{Band structure of manonic crystals}

The semi-analytical approach, relying on plane wave-like solutions, is inherently limited to handling homogeneous magnetization and idealized geometries, such as infinite thin films. In contrast, micromagnetic simulations using the \emph{ms-comsol} module in COMSOL offer far greater flexibility, overcoming these constraints to model complex structures and inhomogeneous (anti-)ferromagnetic textures, with or without dipolar fields. One such scenario is the calculation of spin wave band structures for magnonic crystals.

\begin{figure*}[t]
    \centering
    \includegraphics[width=.98\textwidth]{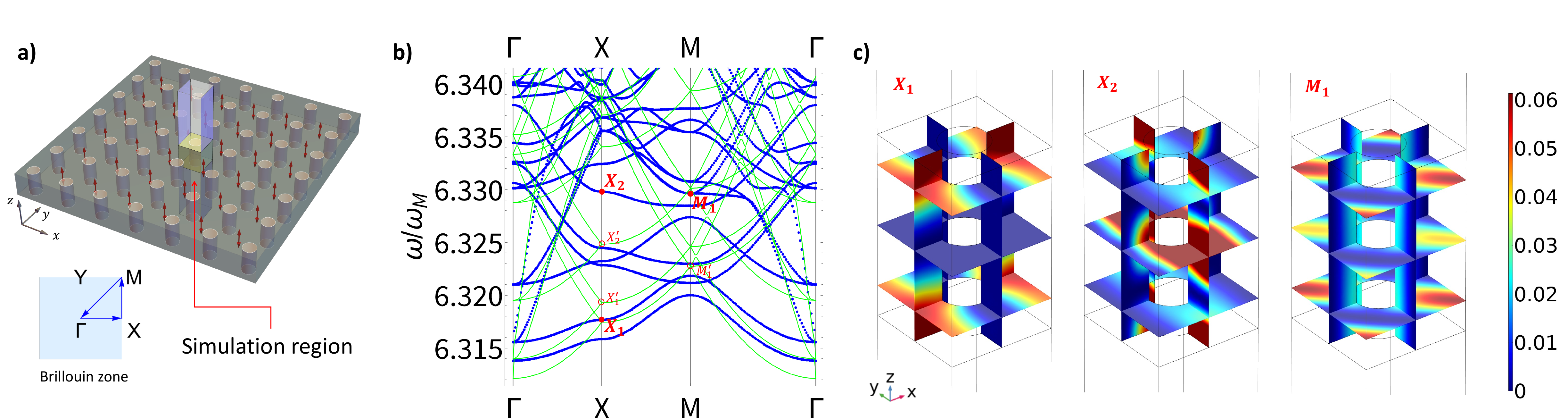}
    \caption{a) The real-space structure of antiferromagnetic magnonic crystals of antidots with cubic lattice. The equilibrium N\'eel vector is perpendicular to the thin film. b) The band structure of anti-dot array along the $\Gamma$-X-M-$\Gamma$ path in the Brillouin zone. The background green lines show the band structure for the complete thin film without holes for comparison. c) The spatial distribution of the net magnetic moment within the unit cell corresponding to points $X_1, X_2$, and $M_1$ on the dispersion curve.}
    \label{fig:crystal}
  \end{figure*}

Traditionally, the calculation of spin wave band structure in magnonic crystals typically require large-scale spatial simulations and extended temporal evolution to obtain adequate data for spatial and temporal Fourier transforms \cite{dvornik_dispersion_2011,kumar_numerical_2011,schwarze_complete_2012,han_wave_2013,krawczyk_review_2014,qin_lowloss_2018}. In contrast, by employing the {\it ms-comsol} method introduced in this paper, the simulation can be confined to a single unit cell in space by using the Floquet periodic boundary condition , similar to the Floquet boundary conditions used in the built-in modules in COMSOL such as the Electromagnetic Waves, Frequency Domain (emw) module. Furthermore, the micromagnetic simulation can be carried out in the frequency domain \cite{zhang_frequency-domain_2023}, thereby obviating the need for time-domain simulations. These simplifications - in both spatial and temporal domains - result in significant reductions in computational time while realizing even better accuracy of the band structure calculations.

We demonstrate the capability to calculate the spin wave band structure of magnonic crystals using a simple example: a periodic array of circular antidots (holes) arranged in a square lattice in an antiferromagnetic thin film. We assume the N\'{e}el vector is perpendicular to the film plane, the radius of the circular holes (antidots) is \( R = \SI{0.4}{\mu m} \), spaced by a periodicity of \( a = \SI{1.8}{\mu m}\), as illustrated in \Figure{fig:crystal}(a). Other parameters are the same as those listed in Table I.  

\Figure{fig:crystal}(b) shows the band structure along $\Gamma$-X-M-$\Gamma$ in the Brillouin zone (The insert of \Figure{fig:crystal}(a)) for the magnonic crystals calculated from {\it ms-comsol}. The band structure for the complete thin film without holes is shown as the green curves in the background for comparison. 
At the boundary of the Brillouin zone, a band gap appears due to the scattering of the corresponding wavelength provided by the holes. 
\Figure{fig:crystal}(c) shows the spatial distribution of the net magnetic moment modulus ($\abs{\mb_1+\mb_2}$) at points $X_1, X_2$ and $M_1$ in the dispersion curves. 
As the radius of the holes decreases to zero, these points continuously trace back to $X'_1, X'_2$ and $M'_1$ on the dispersion for the thin film without holes.
Both $X_1$ and $X_2$ correspond to the first two standing wave in the thickness direction but with the same in-plane wave vector of $(k_x, k_y) = (2\pi/a, 0) $ in the $x$-direction. 
The intensity of $X_1$ is concentrated between the holes, while that of $X_2$ is concentrated between along the holes. 
Additionally, $X_1$ has one node in the thickness direction, while $X_2$ has two, meaning they correspond to the first two standing wave modes in the thickness direction. 
The $M_1$ mode has wave vector $(k_x,k_y) = (2\pi/a, 2\pi/a)$.



Analogous spin wave band structure calculations are feasible for all sorts of (anti-)ferromagnetic magnonic crystals, encompassing those with geometric structures like magnetic antidots or dots of arbitrary shapes and lattice arrangements, and those featuring inhomogeneous periodic magnetic textures such as skyrmion crystals.

\section{Discussion \& Conclusion}

The polarization for the antiferromagnetic spin wave is an interesting topic. Here we consider the polarization defined by the net magnetic moment $\mb = \mb_1 + \mb_2$, or according to the profiles shown in \Figure{fig:disp}. It can be seen from \Figure{fig:disp} that the antiferromagnetic spin wave eigenmodes are linearly polarized for the perpendicular case (the N\'{e}el vector $\bn$ is perpendicular to the film plane) and for the in-plane case with $\bq \parallel \bn$. And for other cases, the eigenmodes are not purely linearly polarized, but have non-trivial elliptical polarization with spatial-dependent ellipticity. There is even possible that the spin wave is right-handed polarized at one location and left-handed at another location.
Some recently developed experimental techniques can be used to detect the polarization of antiferromagnetic spin waves \cite{sheng_control_2025}, and a varying antiferromagnetic spin wave polarization is observed.



We have focused on the dipolar spin waves conventional antiferromagnetic materials. However, approaches described in this paper also apply to the antiferromagnetic van der Waals materials, which have attracted significant attention in recent years. The dipolar-exchange spin waves in these materials are expected to exhibit similar behaviors as the conventional antiferromagnetic materials, but with different material parameter range. Most notbly, because the intralayer and interlayer coupling are quite different in van der Waals materials, the exchange interaction $A, A'$ would become vectors to account different coupling strengths in different directions.

In conclusion, our study has affirmed the validity of micromagnetic simulations for antiferromagnetic spin waves under the influence of dipole fields, utilizing the \textit{ms-comsol} module in COMSOL Multiphysics. By analyzing both dispersion relations and polarization characteristics, we identified distinct behaviors based on the orientation of the N\'eel vector. When the N\'eel vector is perpendicular to the film, the dispersion branches bifurcate into two categories with linear polarization: one with polarization parallel to the film and normal to the wavevector, remaining unaffected by dipole fields, and another with polarization along the wavevector, exhibiting sensitivity to dipole fields for $qd > 1$. Conversely, when the N\'eel vector aligns with the wavevector, the branches again form two linearly polarized modes, where one remains unaffected and parallel to the film, while the other exhibits film-perpendicular polarization influenced by dipole fields even near $q \sim 0$. Interestingly, for the configuration where the N\'eel vector lies parallel to the film but perpendicular to the wavevector, dipole fields impact both polarization modes, leading to their hybridization into elliptically polarized waves. These findings deepen our understanding of antiferromagnetic spin wave dynamics, particularly their interaction with dipole fields under varying geometrical and material constraints.

\bigskip

{\it Acknowledgements.} 
This work was supported by 
National Natural Science Foundation of China (Grants No. 12474110),
the National Key Research and Development Program of China (Grant No. 2022YFA1403300),
the Innovation Program for Quantum Science and Technology (Grant No.2024ZD0300103),
and Shanghai Municipal Science and Technology Major Project (Grant No.2019SHZDZX01).

\bibliographystyle{iopart-num.bst}  
\bibliography{ref,groupx,intro,afm}

\providecommand{\newblock}{}
\begin{thebibliography}{10}
\expandafter\ifx\csname url\endcsname\relax
  \def\url#1{{\tt #1}}\fi
\expandafter\ifx\csname urlprefix\endcsname\relax\def\urlprefix{URL }\fi
\providecommand{\eprint}[2][]{\url{#2}}

\bibitem{gurevichMagnetizationOscillationsWaves1996}
Gurevich A~G and Melkov G~A 1996 {\em Magnetization Oscillations and Waves\/}
  (P: CRC Press) ISBN 0-8493-9460-0 978-0-8493-9460-7

\bibitem{stancilSpinWaves2009}
Stancil D~D and Prabhakar A 2009 {\em Spin {{Waves}}\/} (Office: Springer) ISBN
  0-387-77864-0 978-0-387-77864-8

\bibitem{chumakMagnonSpintronics2015}
Chumak A~V, Vasyuchka V~I, Serga A~A and Hillebrands B 2015 {\em Nature
  Physics\/} {\bf 11} 453--461 ISSN 1745-2473

\bibitem{yu_magnetic_2021}
Yu H, Xiao J and Schultheiss H 2021 {\em Physics Reports\/} {\bf 905} 1--59
  ISSN 03701573

\bibitem{dewames_surface_1969}
De~Wames R~E and Wolfram T 1969 {\em Physical Review\/} {\bf 185} 752--759

\bibitem{goedsche_spin_1970}
Goedsche F 1970 {\em physica status solidi (b)\/} {\bf 41} 711--718 ISSN
  1521-3951

\bibitem{kalinikosSpectrumLinearExcitation1981a}
Kalinikos B~A 1981 {\em Soviet Physics Journal\/} {\bf 24} 718--731 ISSN
  1573-9228

\bibitem{mika_dipolar_1985}
Mika K and Gr{\"u}nberg P 1985 {\em Physical Review B\/} {\bf 31} 4465--4471

\bibitem{kalinikosTheoryDipoleexchangeSpin1986a}
Kalinikos B~A and Slavin A~N 1986 {\em Journal of Physics C: Solid State
  Physics\/} {\bf 19} 7013 ISSN 0022-3719

\bibitem{hillebrands_spinwave_1990}
Hillebrands B 1990 {\em Physical Review B\/} {\bf 41} 530--540

\bibitem{harmsTheoryDipoleexchangeSpin2022}
Harms J~S and Duine R~A 2022 {\em Journal of Magnetism and Magnetic
  Materials\/} {\bf 557} 169426 ISSN 0304-8853

\bibitem{keffer_theory_1952}
Keffer F and Kittel C 1952 {\em Physical Review\/} {\bf 85} 329--337

\bibitem{marshallSpinWaveTheoryAntiferromagnetism1955}
Marshall W 1955 {\em Proceedings of the Royal Society of London. Series A,
  Mathematical and Physical Sciences\/} {\bf 232} 69--77 ISSN 0080-4630
  (\textit{Preprint} \eprint{99683})

\bibitem{oguchi_theory_1960}
Oguchi T 1960 {\em Physical Review\/} {\bf 117} 117--123 ISSN 0031-899X

\bibitem{descloizeauxSpinWaveSpectrumAntiferromagnetic1962}
{des Cloizeaux} J and Pearson J~J 1962 {\em Physical Review\/} {\bf 128}
  2131--2135

\bibitem{loudon_effect_1963}
Loudon R and Pincus P 1963 {\em Physical Review\/} {\bf 132} 673--678

\bibitem{wolfram_surface_1969}
Wolfram T and De~Wames R~E 1969 {\em Physical Review\/} {\bf 185} 762--769

\bibitem{camley_longwavelength_1980}
Camley R~E 1980 {\em Physical Review Letters\/} {\bf 45} 283--286 ISSN
  0031-9007

\bibitem{luthi_dipolar_1983}
L{\"u}thi B and Hock R 1983 {\em Journal of Magnetism and Magnetic Materials\/}
  {\bf 38} 264--268 ISSN 03048853

\bibitem{stamps_dipoleexchange_1987}
Stamps R~L and Camley R~E 1987 {\em Physical Review B\/} {\bf 35} 1919--1931
  ISSN 0163-1829

\bibitem{pereira_theory_1999}
Pereira J~M and Cottam M~G 1999 {\em Journal of Applied Physics\/} {\bf 85}
  4949--4951 ISSN 0021-8979, 1089-7550

\bibitem{wieserQuantizedSpinWaves2009}
Wieser R, Vedmedenko E~Y and Wiesendanger R 2009 {\em Physical Review B\/} {\bf
  79} 144412

\bibitem{shen_magnon_2020}
Shen K 2020 {\em Physical Review Letters\/} {\bf 124} 077201 ISSN 0031-9007,
  1079-7114

\bibitem{liuDipolarSpinWaves2020}
Liu J, Wang L and Shen K 2020 {\em Physical Review Research\/} {\bf 2} 023282

\bibitem{sun_dipolar_2024}
Sun Y, Meng F, Lee C, Soll A, Zhang H, Ramesh R, Yao J, Sofer Z and Orenstein J
  2024 {\em Nature Physics\/} {\bf 20} 794--800 ISSN 1745-2481

\bibitem{chengAntiferromagneticSpinWave2016}
Cheng R, Daniels M~W, Zhu J~G and Xiao D 2016 {\em Scientific Reports\/} {\bf
  6} 24223 ISSN 2045-2322

\bibitem{lan_antiferromagnetic_2017}
Lan J, Yu W and Xiao J 2017 {\em Nature Communications\/} {\bf 8} 178 ISSN
  2041-1723

\bibitem{proskurinSpinWaveChiralityIts2017}
Proskurin I, Stamps R~L, Ovchinnikov A~S and Kishine J~i 2017 {\em Physical
  Review Letters\/} {\bf 119} 177202

\bibitem{yu_polarization-selective_2018}
Yu W, Lan J and Xiao J 2018 {\em Physical Review B\/} {\bf 98} 144422
  \urlprefix\url{https://link.aps.org/doi/10.1103/PhysRevB.98.144422}

\bibitem{abertMicromagneticsSpintronicsModels2019}
Abert C 2019 {\em The European Physical Journal B\/} {\bf 92} 120 ISSN
  1434-6036

\bibitem{_comsol_}
Comsol multiphysics COMSOL AB

\bibitem{yu_comsol_nodate}
Yu W {COMSOL} {Blog}: {Micromagnetic} {Simulation} with {COMSOL} {Multiphysics}
  \urlprefix\url{https://www.comsol.com/blogs/micromagnetic-simulation-with-comsol-multiphysics/}

\bibitem{solovyev_magnetooptical_1997}
Solovyev I~V 1997 {\em Physical Review B\/} {\bf 55} 8060--8063 ISSN 0163-1829,
  1095-3795

\bibitem{treves_magnetic_1962}
Treves D 1962 {\em Physical Review\/} {\bf 125} 1843--1853

\bibitem{lan_spin-wave_2015}
Lan J, Yu W, Wu R and Xiao J 2015 {\em Physical Review X\/} {\bf 5} 041049
  \urlprefix\url{http://link.aps.org/doi/10.1103/PhysRevX.5.041049}

\bibitem{yu_magnetic_2016}
Yu W, Lan J, Wu R and Xiao J 2016 {\em Physical Review B\/} {\bf 94} 140410
  \urlprefix\url{http://link.aps.org/doi/10.1103/PhysRevB.94.140410}

\bibitem{yu_magnetic_2020}
Yu W, Lan J and Xiao J 2020 {\em Physical Review Applied\/} {\bf 13} 024055
  \urlprefix\url{https://link.aps.org/doi/10.1103/PhysRevApplied.13.024055}

\bibitem{yu_hopfield_2021}
Yu W, Xiao J and Bauer G~E~W 2021 {\em Physical Review B\/} {\bf 104} L180405
  publisher: American Physical Society
  \urlprefix\url{https://link.aps.org/doi/10.1103/PhysRevB.104.L180405}

\bibitem{zhang_frequency-domain_2023}
Zhang J, Yu W, Chen X and Xiao J 2023 {\em AIP Advances\/} {\bf 13} 055108 ISSN
  2158-3226 \urlprefix\url{https://doi.org/10.1063/5.0143262}

\bibitem{vandenboom_fully_2021}
{van den Boom} S~J, {van Keulen} F and Arag{\'o}n A~M 2021 {\em Computer
  Methods in Applied Mechanics and Engineering\/} {\bf 382} 113848 ISSN
  0045-7825

\bibitem{yan_allmagnonic_2011}
Yan P, Wang X~S and Wang X~R 2011 {\em Physical Review Letters\/} {\bf 107}
  177207 ISSN 0031-9007, 1079-7114

\bibitem{prabhakar_spin_2009}
Prabhakar A and Stancil D~D 2009 {\em Spin Waves: Theory and Applications\/}
  (Boston, MA: Springer US) ISBN 978-0-387-77864-8 978-0-387-77865-5

\bibitem{dvornik_dispersion_2011}
Dvornik M and Kruglyak V~V 2011 {\em Physical Review B\/} {\bf 84} 140405 ISSN
  1098-0121, 1550-235X

\bibitem{kumar_numerical_2011}
Kumar D, Dmytriiev O, Ponraj S and Barman A 2011 {\em Journal of Physics D:
  Applied Physics\/} {\bf 45} 015001 ISSN 0022-3727

\bibitem{schwarze_complete_2012}
Schwarze T, Huber R, Duerr G and Grundler D 2012 {\em Physical Review B\/} {\bf
  85} 134448 ISSN 1098-0121, 1550-235X

\bibitem{han_wave_2013}
Han D~S, Vogel A, Jung H, Lee K~S, Weigand M, Stoll H, Sch{\"u}tz G, Fischer P,
  Meier G and Kim S~K 2013 {\em Scientific Reports\/} {\bf 3} 2262 ISSN
  2045-2322

\bibitem{krawczyk_review_2014}
Krawczyk M and Grundler D 2014 {\em Journal of Physics: Condensed Matter\/}
  {\bf 26} 123202 ISSN 0953-8984

\bibitem{qin_lowloss_2018}
Qin H, Both G~J, H{\"a}m{\"a}l{\"a}inen S~J, Yao L and {van Dijken} S 2018 {\em
  Nature Communications\/} {\bf 9} 5445 ISSN 2041-1723

\bibitem{sheng_control_2025}
Sheng L, Duvakina A, Wang H, Yamamoto K, Yuan R, Wang J, Chen P, He W, Yu K,
  Zhang Y, Chen J, Hu J, Song W, Liu S, Han X, Yu D, Ansermet J~P, Maekawa S,
  Grundler D and Yu H 2025 {\em Nature Physics\/}  1--6 ISSN 1745-2481

\bibitem{dewames_dipoleexchange_1970}
De~Wames R~E and Wolfram T 1970 {\em Journal of Applied Physics\/} {\bf 41}
  987--993 ISSN 0021-8979

\end{thebibliography}

\newpage
\appendix 

\section{Exchange coupling in AFM}
\label{app:exchange}

The energy form of the lattice-regularized antiferromagnetic exchange interaction is:
\begin{equation}
  E_A = J_A\bS_1\cdot\bS_2
\end{equation}
When the spins of adjacent atoms are arranged in the opposite direction, the system is in a low-energy state. The dimension of $J_A$ is \si{\mu_0M_S^2V}.
Consider an alternating atomic structure: $a_0, b_0, a_1, b_1, a_2, b_2, a_3, b_3, \ldots$ For an atom on the $a$ sublattice, the exchange interaction energy is:
\begin{equation}
  E_A = J_A\bS_{ai}\cdot(\bS_{b(i-1)} + \bS_{bi})
\end{equation}
The process of continuous approximation can be regarded as spreading the magnetic moments concentrated on the atoms over the unit cell. In this process, the dimensions of spatial volume and $M_s$ are removed, resulting in a dimensionless $\mb(x)$:
\begin{equation}
  \begin{aligned}
      e_A &= \frac{nJ_A}{(2a)^3}\mb_a\cdot\left(2\mb_b + a^2\nabla^2\mb_b\right) \\
      &= \frac{nJ_A}{4a^3}\mb_a\cdot\mb_b - \frac{nJ_A}{8a^3}a^2\mathbf{\nabla m_a}\cdot\mathbf{\nabla m_b}
  \end{aligned}
\end{equation}
The latter term is expanded in a Taylor series, where $a$ is the unit cell edge length, and $n$ is the number of atoms with antiferromagnetic interactions with a single atom. The parameters are concentrated into two terms: direct coupling between different lattices and exchange interaction:
\begin{equation}
  e_A = M_s(J\mb_a\cdot\mb_b + A\mb_a\cdot\mathbf{\nabla^2m_b})
\end{equation}
where
\begin{equation}
      J = \frac{n_1J_A}{4M_sa^3} \qand
      A = \frac{n_1J_A}{8M_sa}
\end{equation}
Systems that can be described by the LLG equation in a continuous manner require that $a \ll \lambda$, where $\lambda$ is the wavelength of the spin wave, \ie:
\begin{equation}
  \frac{A}{J} \ll \lambda^2 \sim \frac{1}{q^2 + q_x^2}
\end{equation}
where $q$ is the in-plane wave vector of the antiferromagnetic film, and $q_x$ is the wave vector perpendicular to the film (with the film normal along the $x$ direction).

Using the same approach, the expression for the ferromagnetic exchange interaction can be derived as
\begin{equation}
    e_{Fi} = -\half M_sA'\mb_i\cdot\nabla^2\mb_i \qwith
    A' = \frac{n_2J_F}{8M_sa'}
\end{equation}
where $a'$ is the distance within the same ferromagnetic lattice, $J_F$ is the ferromagnetic exchange coupling, and $n_2$ is the number of atoms that have ferromagnetic exchange coupling with a specific atom.
In order to match the form of antiferromagnetic exchange, we define:
\begin{equation}
  J' = \frac{n_2J_F}{4M_sa'^3}
\end{equation}

\section{Semi-analytical Approach}
\label{app:semi-analytical}

For antiferromagnetic thin film with uniform N\'eel vector case, a semi-analytical approach is possible by mimicing the calculation of the dipolar-exchange spin wave for ferromagnetic thin film by De Wames and Wolfram \cite{dewames_dipoleexchange_1970}, and for antiferromagnet by Stamps and Camley \cite{stamps_dipoleexchange_1987}. 

The coupled LLG equations \Eq{eqn:LLGmm} can be converted into the equations of motion for the N\'{e}el vector $\bn = \mb_1 - \mb_2$ ($\abs{\bn} \simeq 2$) and net magnetization $\mb = \mb_1 + \mb_2$ ($ \abs{\mb}\ll 1, \mb \cdot \bn = 0$). Keeping up to the first order in small quantity of $\mb, \bh$ and neglecting the damping, \Eq{eqn:LLGmm} is rewritten as:
\begin{subequations}
  \label{eqn:LLGmn}
  \begin{align}
    \frac{\dot{\bn}}{\gamma}
    &= - \mb\times K\hat{\bz} 
    -\bn\times\qty(\bH_\text{ext} - J \mb + \bh - \frac{A-A'}{2}\nabla^2 \mb) \nn 
    &\simeq - \mb\times K\hat{\bz} -\bn\times\qty(\bH_\text{ext} - J \mb + \bh) \\
    \frac{\dot{\mb}}{\gamma}
    &= -\mb\times \bH_\text{ext} 
    - \bn\times\qty(K\hat{\bz} + \frac{A + A'}{2}\nabla^2 \bn).
  \end{align}
\end{subequations}
The approximation in the equation for $\dot{\bn}$ assumes that $J \gg (A-A')\nabla^2$, \ie $J \gg (A-A')q^2$, thus the $\nabla^2$ term in the $\dot{\bn}$ equation can be neglected. 
With this approximation, $A$ and $A'$ only appear in the EOM for $\dot{\mb}$, and have exactly the same effect on the results, therefore there is no need to distinguish $A, A'$. In the simulation, we can replace the inter-lattice exchange interaction $A$ with the exchange interaction $A'$ within a single lattice.
From \Eq{eqn:LLGmn}, one see that the dipolar field only enters the EOM for $\dot{\bn}$. The effect of dipolar field on $\dot{\mb}$ has the form of $\mb\times\bh(\mb)$, which is second order in $\mb$, thus neglected in the linearized equation above.

We now consider the equilibrium magnetic moments of the two sublattices point in $\hbz$: $\mb_1^0 = \hbz, \mb_2^0 = -\hbz$. And let $\mb_\perp = (m_x,m_y)$, $\bn_\perp = (n_x,n_y)$ be the small deviation due to excitation, and $\bh_\perp = (h_x,h_y)$ is the resulting dipolar fields due to $\mb_\perp$. Note that $\mb_z \sim 0$, $\bn_z \sim 2$, \Eq{eqn:LLGmn} can be rewritten as a linear response equation
\begin{equation}
  \label{eqn:LLGmn2}
  \mb_{\perp} = \chi_m~ \gamma\bh_{\perp} \qand 
  \bn_{\perp} = \chi_n~ \gamma\bh_{\perp},
\end{equation}
where the magnetic and N\'{e}el susceptibility ($\omega_0 \equiv \gamma H_\text{ext}$)
\begin{subequations}
\label{eqn:chi}
\begin{align}
    \chi_m &= \frac{\hat{2\Omega}}{\hW^4 - 4\omega^2\omega_0^2} 
    \mqty( \hW^2 & -2i\omega\omega_0 \\ 2i\omega\omega_0 & \hW^2) 
    \overset{\omega_0=0}{\lra} 2\hat{\Omega}\hat{W}^{-2}, \\
    \chi_n &=
    \mqty( -\omega_0 & i\omega \\ -i\omega & -\omega_0 ) \hat{\Omega}^{-1} \chi_m 
    \overset{\omega_0=0}{\lra} -\omega\sigma_y\hat{W}^{-2},
\end{align}
\end{subequations}
where we defined two operators
\begin{equation*}
\hW^2 \equiv (K + 2J)\hat{\Omega} - \omega^2 - \omega_0^2
\qand 
\hat{\Omega} \equiv K - (A + A')\nabla^2. 
\end{equation*}
The second equal signs in \Eq{eqn:chi} applies in the absence of external field ($\omega_0 = 0$).

In quasistatic approximation, the dipolar magnetic field $\bh(\br, t)$ satisfies $\nabla\times\bh = 0$, 
which means that the field can be represented by a scalar potential: $\bh(\br, t) = -\nabla\psi(\br, t)$.
The magnetic flux $\bb(\br, t)$ satisfy the Maxwell's equations in the quasistatic approximation:
\begin{subequations}
  \label{eqn:maxwell}
  \begin{align}
    \nabla\cdot\bb = \nabla\cdot(\bh + \mu_0M_s\mb) &= 0 \quad \text{inside the film,} \\
    \nabla\cdot\bb = \nabla\cdot\bh &= 0 \quad \text{outside the film,}
  \end{align}
\end{subequations}
The transverse dipolar field is related to the full dipolar field as $\bh_\perp = \bh - \hat{\bz}h_z = -\nabla\psi + \hat{\bz}\p_z\psi$. 
\Eqss{eqn:LLGmn2}{eqn:maxwell} completely determine the properties of the dipolar-exchange spin waves in antiferromagnet.

The net magnetization $\mb$ can be written in terms of the scalar potential $\psi$ via $\mb \simeq \mb_\perp = \chi_m\gamma\bh_\perp$, therefore the equations in \Eq{eqn:maxwell} can be also be expressed in terms of the scalar potential ($\omega_\ssf{M} \equiv \gamma M_s$):
\begin{subequations}
  \label{eqn:psi}
  \begin{align}
    \qty[(\hW^2 + 2\omega_\ssf{M}\hat{\Omega})\hW^2 - 4\omega^2\omega_0^2]\nabla^2\psi &= 2\omega_\ssf{M}\hat{\Omega}\hW^2\p_z^2\psi, \label{eqn:psi_in} \\ 
    \nabla^2\psi &= 0. \label{eqn:psi_out}
  \end{align}
\end{subequations}
The equations above, equivalent to \Eqss{eqn:LLGmn2}{eqn:maxwell}, governs the full dynamical information about the dipolar fields and the magnetic moments of the antiferromagnetic thin film. Once the scalar potential $\psi$ is solved using \Eq{eqn:psi}, the dipolar field and the magnetic moment $\mb(\br,t)$ can be infered straightforwardly.
Since both $\hW^2$ and $\hat{\Omega}$ involve Laplacian operators, \Eq{eqn:psi} is a set of 6th-order ordinary differential equations. If restore the neglected term in \Eq{eqn:LLGmn}, the 6th-order becomes 10th-order. In comparison, the equivalent equation for ferromagnetic case is 6-th order, too \cite{dewames_dipoleexchange_1970}.

The EOM for $\psi(\br,t)$ in \Eq{eqn:psi} can only be solved with proper boundary conditions at the interfaces between the antiferromagnetic thin film and the vacuum. We use another coordinates $(x',y',z')$ with $z'$ perpendicular to the thin film plane, and the two interfaces are at $z' = 0, d$. The easy axis $\hbz =  \hbx'\sin\eta + \hbz' \cos\eta$, or the film normal $\hbz' = \hbx \sin\eta + \hbz \cos\eta$.
The boundary conditions for the dipolar fields require the continuity of the following components across the interfaces:
\begin{equation}
  \label{eqn:bc_hb}
  \left.\hbz'\times \bh\right|_{z'=0, d},
  \qand
  \left.\hbz'\cdot \bb\right|_{z'=0, d}.
\end{equation}
In addition, by integrating the second equation of \Eq{eqn:LLGmn} over a tiny distance across the interfaces, we find another set of boundary condition, \ie zero flux for $\bn$:
\begin{equation}
  \label{eqn:bc_n}
  \left.\bn\times\pdv{\bn}{z'}\right|_{z' = 0,d} = 0.
\end{equation}



Let $\bu = (x', y')$ and $\bk = (k_{x'}, k_{y'})$ be the in-plane position and wavevector, 
we use following ansatz solution in \Eq{eqn:psi}:
\begin{equation}
  \psi(\br,t) = 
  \begin{cases}
      \phi(0)e^{+kz'}e^{-i\bk\cdot\bu}e^{i\omega t} & \quad z' < 0 \\
      \phi(z')e^{-i\bk\cdot\bu}e^{i\omega t} & \quad 0 < z' < d \\
      \phi(d)e^{-kz'}e^{-i\bk\cdot\bu}e^{i\omega t} & \quad z' > d
  \end{cases}.
\end{equation}
This ansatz enables the replacement of the Laplacian operator $\nabla^2\ra \partial_{z'}^2 - k^2$, and it automatically satisfies \Eq{eqn:psi_out} as well as the boundary conditions for $\bh$ in \Eq{eqn:bc_hb}. 
The remaining equation need to be solved is \Eq{eqn:psi_in} with the zero-flux boundary condition \Eq{eqn:bc_n} and the boundary conditions for $\bb$ in \Eq{eqn:bc_hb}.
\Eq{eqn:psi_in} is a standard six-order ordinary differential equation of $\phi(z')$ in $\partial_{z'}$,
which can be solved via the Euler method with six exponential solutions:
\begin{equation}
  \label{eqn:phi}
  \phi(z') 
  = \sum_{j = 1}^6c_je^{-iq_jz'}
  = \sum_{j = 1}^3(a_je^{iq_jz'} + b_je^{-iq_jz'}),
\end{equation}
where $\pm q_j$ are the six roots to the polynomial equation corresponding to the 6-th ordinary differential equation \Eq{eqn:psi_in}.
The six exponential are rearranged into positive and negative pairs because
\Eq{eqn:psi_in} only contains the even order of $\partial_{z'}$.
The six unknown coefficients $a_j, b_j$ are determined by the six boundary conditions for $\mb$ in \Eq{eqn:bc_hb} and for zero-flux in \Eq{eqn:bc_n} at $z' = 0, d$ :
\begin{subequations}
  \label{eqn:BC_phi}
  \begin{align}
    \p_{z'}\phi + \sin\eta \qty(\chi_m^{xx}\p_x - i\chi_m^{xy}k_y)\phi \mp k\phi &= 0, \\
    \p_{z'}
    \chi_n \mqty(\p_x \\ -ik_y) \phi &= 0.
  \end{align}
\end{subequations}
where $\p_{z'} = \sin\eta \p_x + \cos\eta \p_z$, and $\pm k$ in the first equation corresponds to the two interfaces at $z' = 0, d$, respectively.
By plugging \Eq{eqn:phi} in the equations above, \Eq{eqn:BC_phi} becomes a set of linear equations with six unknowns:
\begin{equation}
  \label{eqn:Mab}
  M(\eta,\omega,\bk) \mqty(a_1 & b_1 & a_2 & b_2 & a_3 & b_3)^T = 0,
\end{equation}
The dispersion of the dipolar-exchange antiferromagnetic spin wave for a given in-plane wave-vector $\bk$ is solved by finding the roots in $\omega$ that making the determinant of the coefficient matrix vanish:
\begin{equation}
  \label{eqn:Mwk}
  \abs{M(\eta, \omega, \bk)} = 0.
\end{equation}
The corresponding spin wave profile is found by solving for the coefficients $a_j, b_j$ in \Eq{eqn:Mab} at the resonant frequencies.

\section{Dipole surface spin waves in the long-wavelength limit}
\label{app:surface_wave}

For the in-plane magnetized case ($\bn_0 \perp \hbz'$), when the in-plane wave vector is perpendicular to the in-plane N\'{e}el vector ($\bk\perp \bn_0$), the equation of motion for the magnetic scalar potential $\psi$ in the long wavelength limit ($qd \ll 1$) is:
\begin{equation}
    \qty[(\hW^2 + 2\omega_\ssf{M}K)\hW^2 - 4\omega^2\omega_0^2](k^2 + q_j^2)\psi = 0, 
\end{equation}
which is solved by 
\begin{equation}
  k^2 + q_j^2 = 0 \qRa q_j = \pm i k.
\end{equation}
This solution corresponds to a surface spin wave. By substituting $q_j = \pm ik$ into the boundary condition given by \Eq{eqn:BC_phi}:
\begin{equation}
  \qty(1 + \frac{2\omega_MK\hW^2}{\hW^4 - 4\omega^2\omega_0^2})\p_x\phi + \frac{4\omega\omega_0\omega_MK}{\hW^4 - 4\omega^2\omega_0^2}k\phi \mp k\phi = 0,
\end{equation}
we have the dispersion for the surface spin wave:
\begin{equation}
  \begin{aligned}
    \omega^2 &= K(K + 2J + \omega_\ssf{M}) + \omega_0^2 \\
    &\quad \pm \sqrt{4K(K + 2J + \omega_\ssf{M})\omega_0^2 + \omega_\ssf{M}^2K^2e^{-2kd}}.
  \end{aligned} 
\end{equation}
They are the analytical expressions for the lowest-order modes of the two dispersion branches in \Figure{fig:disp}c when $qd \ll 1$, exhibiting surface wave characteristics. In these two modes, the excitation strengths of the magnetic moments of the two sublattices are concentrated on the upper and lower surfaces, respectively. However, the total magnetic moment does not possess surface wave characteristics. Only when an external magnetic field is applied can a distinct surface wave characteristic be observed in the net magnetic moment.

\end{document}